# Growth, structure analysis and anisotropic superconducting properties of MgB$_2$ single crystals.

S. Lee, H. Mori, T. Masui, Yu. Eltsev, A. Yamamoto and S. Tajima[*]

Superconductivity Research Laboratory, ISTEC, 1-10-13 Shinonome, Koto-ku, Tokyo 135-0062, Japan

Here we report the growth of sub-millimeter MgB$_2$ single crystals of various shapes under high pressure in Mg-B-N system. Structure refinement using a single-crystal X-ray diffraction analysis gives lattice parameters $a$=3.0851(5)Å and $c$=3.5201(5)Å with small reliability factors ($R_w$=0.025, $R$=0.018), which enables us to analyze the Fourier and Fourier difference maps. The maps clearly show the B $sp^2$ orbitals and covalency of the B-B bonds. The sharp superconducting transitions at $T_c$=38.1-38.3K were obtained in both magnetization ($\Delta T_c$=0.6K) and resistivity ($\Delta T_c$<0.3K) measurements. Resistivity measurements with magnetic fields applied parallel and perpendicular to the Mg and B sheets reveal the anisotropic nature of this compound, with upper critical field anisotropy ratio of about 2.7.

The discovery of superconductivity at 39K in magnesium diboride (MgB$_2$)[1] has led to intense interest in the physics, chemistry and material science of this compound[2-5]. At present, most of the experiments were performed on polycrystalline samples where random orientation of crystallites and presence of the impurity phases resulted in a poor reproducibility of the reported data. Obviously, high-quality single crystals are indispensable for further studies. Owing to the thermal instability of MgB$_2$, crystal growth at ambient pressure is impossible. In the case of heat treatment in sealed quartz and metal (Ta, Nb, Mo) tubes, the temperature range is limited and contamination from container materials can not be completely avoided which results in small or impure crystals. High-pressure technique could be very useful to prevent the evaporation of Mg from the reaction mixture and to suppress the decomposition of MgB$_2$. Recently, effect of sintering temperature 500-950°C at 3 GPa on the superconducting properties of MgB$_2$ was studied[6]. It was found that as sintering temperature increased, the sample quality was improved, but for the heat treatment above 950°C the gold started to melt and to adhere strongly to MgB$_2$. Preparation of high-density bulk MgB$_2$ materials at a higher temperature (950-1250°C) and 3.5 GPa was reported in BN crucibles[7]. However, the samples prepared at 1250°C apparently showed inhomogeneous structure in macroscopic scale, indicating the occurrence of contamination from the crucible wall. These results are not quite favorable for crystal growth of MgB$_2$ under high-pressure.

In this letter, we report the successful growth of high-quality MgB$_2$ single crystals under high pressure in the Mg-B-N system. Using the crystals, we carried out structure refinement and measured anisotropic superconducting properties of magnesium diboride.

We have grown single crystals of $MgB_2$ in the quasi-ternary $Mg$-$MgB_2$-$BN$ system at a pressure of 4-6 GPa and temperature 1400-1700°C for 5 to 60 min. The average weight ratio of $Mg:MgB_2:BN$ was 1:3:15. The $MgB_2$ precursor was synthesized from magnesium powder (99.9% Rare Metallic Co.) and amorphous boron (97% Hermann C.Starck). The experiments were performed in BN containers using a cubic-anvil press (TRY Engineering). The longitudinal temperature gradient between the center and outer parts of container was approximately 200°C/cm.

Diffraction data were collected at room temperature by a $w$–$2q$ scan, using a Rigaku AFC5R four-circle diffractometer (50 kV, 150mA) with $MoK_a$ radiation monochromized by graphite ($l$=0.71069Å). Among the measured 976 reflections, 96 were used for refinement after absorption correction by a $y$-scan. The structure was refined by the full-matrix least-square procedure.

The temperature dependence of the magnetization $M(T)$ was measured by a SQUID (MPMS XL, Quantum Design) for 37 crystals of different shape (total weight 0.25mg) aligned with the $c$-axis perpendicular to the sample holder. In-plane electrical resistivity was measured for several plate-like crystals of typical size of 500x100x20 µm$^3$ in a four-probe configuration using a low frequency (17.8Hz) ac technique with a voltage resolution of less than 0.3nV. To study the anisotropic superconducting properties, the resistivity measurements were performed in the magnetic fields up to 7 Tesla applied parallel and perpendicular to the Mg and B planes.

The Mg-B-N system has been extensively studied during last 40 years[8-10]. In this system Mg and Mg-containing compounds act as the catalysts for the transformation of boron nitride (BN) from hexagonal to cubic form under high pressure and the formation of $MgB_2$ and $MgB_6$ crystals has been observed[11]. These compounds were formed as the intermediate phases in Mg-BN system, during the synthesis of cubic boron nitride, but $MgB_2$ was decomposed after prolong heat treatment.

We found that the lower limit for $MgB_2$ crystal growth in "pressure-temperature" space is located above the line of eutectic liquid formation in the Mg-B-N system while the upper limit is the decomposition line of $MgB_2$ to $MgB_4$. The location of both lines strongly depends on the presence of impurity phases ($MgB_4$, MgO, $B_2O_3$) in the starting materials. In optimal conditions shiny yellow-colored single crystals with size of 200-700 μm were grown. The size and shape of the crystals are governed by the temperature, duration of heat treatment and Mg/B ratio and varied from needle-like to hexagonal thick plates. Figure 1 shows a scanning electron microscopy image of the single crystals extracted mechanically from a bulk sample.

Although crystal structure of $MgB_2$ was established in 1954[12], using polycrystalline and powdered sample, refined structural data were previously unavailable. It has been proved to be very difficult to avoid the formation of impurity phases (MgO and $MgB_4$) in polycrystalline samples and thus the precise structure refinement is almost impossible. In this study well-formed single crystals were chosen for structure analysis to minimize of absorption effects. The refined crystallographic data, atomic coordinates, and thermal parameters are listed in Table 1. The small values of reliability factors ($R_w$=0.025, $R$=0.018) enable us to analyze the Fourier and Fourier difference maps. The Fourier and Fourier difference maps in the B-plane are presented in Figs. 2(a) and (b), respectively. The B atom clearly shows the electron distribution of a $sp^2$ orbital in Fig.2(a) while unambiguous covalent bonds between B atoms form the hexagonal network in the Fourier difference map of Fig.2(b). In contrast, there is no bond electron between Mg and B atoms, which suggests no covalency between them. These results are quite consistent with the band calculation that predicts a predominant contribution of B *p*-orbitals to carrier conduction[13].

In the final part of this letter we briefly characterize the superconducting properties of our MgB$_2$ single crystals. Figure 3 shows the *M(T)* curves in the zero-field cooling (ZFC) and field-cooling (FC) modes. Onset of superconducting transition was observed at $T_c$=38.1K with a transition width ***D**T$_c$*(10-90%)=0.6K indicating the high quality and homogeneity of the samples.

Figure. 4 shows a typical temperature dependence of in-plane resistivity at zero field. We found sharp superconducting transitions around 38.1-38.3K with ***D**T*$_c$(10-90%)=0.2-0.3K. Note that we performed several independent measurements of resistivity on various MgB$_2$ crystals and got a very high reproducibility, which proves that our crystals are of high quality and all obtained results are reliable. The estimated resistivity at 40K is about 1 µΩcm and the residual resistivity ratio (RRR) ***r***(273K)/***r***(40K)=5±0.1. Our results settle down the problem of scattered resistivity ranging from 0.38 to 360 µΩcm at 40K among the reports on different samples[4,14,15].

In Fig.5(a) we present the resistive superconducting transitions at zero magnetic field and under magnetic field 2T, applied parallel (*H//*) and perpendicular (*H*⊥) to the Mg and B planes. One can see a rapid suppression of superconductivity for the perpendicular field compared to the parallel one. The temperature dependence of upper critical field $H_{c2}$ at $H \perp ab$ and $H//ab$ is shown in Fig.5(b). The anisotropy ratio is estimated as $h=H_{c2//}/H_{c2\perp}=2.7$ that is slightly larger than the reported value $h$=1.6-2.0 for thin films and aligned crystallites[14,15]. In spite of rather small anisotropy of $MgB_2$ in comparison with the superconducting cuprates, it strongly affects the irreversible properties of this material. The remarkable broadening of the superconducting transition observed for $H\perp ab$ indicates a significant suppression of the irreversibility field $H^*\perp$ below $H_{c2\perp}$, giving $H^*\perp(T)\approx 0.5H_{c2\perp}(T)$. In sharp contrast, in the parallel field orientation $H^*_{//}(T)\approx 0.93H_{c2//}(T)$.

In conclusion, we have successfully developed the high-pressure technique for $MgB_2$ single crystals growth in the quasi-ternary Mg-BN-$MgB_2$ system. For these crystals, we have carried out a precise structure analysis using a four-circle X-ray diffraction measurement, obtained the full structure parameters and analyzed the electron density maps (EDM). Results of EDM clearly demonstrate a covalent nature of the B-B bonding, which is in good agreement with the theoretical prediction. In addition to this orbital information, we have obtained an appreciable anisotropy in the superconducting properties.

## Acknowledgements

This work was supported by the New Energy and Industrial Technology Development Organization (NEDO) as Collaborative Research and Development of Fundamental Technologies for Superconductivity Applications.

# References

1) J.Nagamatsu, N.Nakagawa, T.Muranaka, Y.Zentani and J.Akimitsu: Nature **410** (2001) 63.

2) D.C.Larbalestier, M.Rikkel, L.D.Cooley, A.A.Polyanskii, J.Y.Jiang, S.Patnaik, X.Y.Cai, D.M.Feldman, A.Gurevich, A.A. Squitieri, M.T.Naus, C.B.Eom, E.E.Hellstrom, R.J.Cava, K.A.Regan, N.Rogado, M.A.Hayward, T.He, J.S.Slusky, P.Khalifah, K.Inumatu and M.Haas: Nature **410** (2001) 186.

3) J.S.Slusky, N.Rogado, K.A.Regan, M.A.Hayward, P.Khalifah, T.He, K.Inumaru, S.Loureiro, M.K.Haas, H.W.Zandbergen and R.J.Cava: Nature **410** (2001) 186.

4) D.K.Finnemore, J.E.Ostenson, S.L.Bud'ko, G.Lapertot and P.C.Canfield: Phys.Rev.Lett. **86** (2001) 2420.

5) P.C.Canfield, D.K.Finnemore, S.L.Bud'ko, J.E.Ostenson, G.Lapertot, C.E.Cunningham and C.Petrovic: Phys.Rev.Lett. **86** (2001) 2423.

6) C.U.Jung, M.S.Park, W.N.Kang, M.S.Kim, K.H.P.Kim, S.Y.Lee and S.I.Lee: Physica C **353** (2001) 162.

7) Y.Takano, H.Takeya, H.Fujii, H.Kumakura, T.Hatano, K.Togano, H.Kito and H.Ihara: Appl.Phys.Lett. **78** (2001) 2914.

8) R.H.Wentorf Jr.: J.Chem.Phys. **34** (1961) 809.

9) T.Endo, O.Fukunaga, M.Iwata: J.Mater.Sci. **14** (1979) 1676.

10) V.L.Solozhenko, V.Z.Turkevich, W.B.Holzapfel: J.Phys.Chem.B **103** (1999) 8137.

11) N.E.Filonenko, V.I. Ivanov, L.I. Fel'dgun, M.I. Sokhor and L.F. Vereshchagin: Dokl. Akad. Nauk SSSR **175** (1967) 833.

12) M.E.Jones and R.E.Marsh: J.Am.Chem.Soc. **76** (1954) 1434.

13) J.Kortus, I.I.Mazin, K.D.Belashchenko, V.P.Antropov and L.L.Boyer: Phys.Rev.Lett. **86** (2001) 4656.

14) S.Patnaik, L.D.Cooley, A.Gurevich, A.A.Polyanskii, J.Jiang, X.Y.Cai, A.A.Squitieri, M.T.Naus, M.K.Lee, J.H.Choi, L.Belenky, S.D.Bu, J.Letter, X.Song, D.G.Schlom, S.E.Babcock, C.B.Eom, E.E.Hellstrom and D.C.Larbalestier: Supercond.Sci.Technol. **14** (2001) 315.

15) O.F. deLima, R.A.Ribeiro, M.A.Avila, C.A.Cardoso and A.A.Coelho: to be published in Phys.Rev.Lett. **86** (2001).

# Figure captions

Figure 1. Morphology of single crystals of $MgB_2$.

Figure 2. (a) Fourier map for $MgB_2$ in the B-plane; (b) Fourier difference map for $MgB_2$ in the B-plane. Contour intervals are 0.3 and 0.05 e/Å$^3$, respectively. Contours in positive (negative) regions are represented by solid (dashed) curves.

Figure 3. Temperature dependence of the magnetization after zero field cooling (ZFC) and field cooling (FC) in a magnetic field $\mu_0 H$=1mT parallel to the *c*-axis of 37 aligned single crystals. Inset: expanded view of temperature dependence of the magnetization near $T_c$.

Figure 4. Temperature dependence of the in-plane resistivity at zero field. Inset: expanded view for temperature dependence of the in-plane resistivity near $T_c$.

Figure 5. (a) Temperature dependence of the in-plane resistivity at zero magnetic field and under magnetic field 2T for $H \perp ab$ and $H//ab$. The arrows show how $H_{c2}$ and $H^*$ were determined; (b) The upper critical field as the function of temperature at $H \perp ab$ and $H//ab$.

Table 1  Crystallographic data, atomic coordinates, and thermal parameters for $MgB_2$

| Atom | Site | Atomic coordinates | | | Thermal parameters (x $10^4$ Å$^2$) | | | | |
| --- | --- | --- | --- | --- | --- | --- | --- | --- | --- |
| | | $x$ | $y$ | $z$ | $U_{11}$ | $U_{22}$ | $U_{33}$ | $U_{12}$ | $B_{eq.}$ (Å$^2$) |
| Mg | a | 0 | 0 | 0 | 48 | 48(2) | 56(3) | 24 | 0.399(7) |
| B | d | 1/3 | 2/3 | 1/2 | 47 | 47(3) | 51(5) | 23 | 0.38(1) |

Hexagonal: space group *P6/mmm* (no191), *a*=3.0851(5)Å, *c*=3.5201(5)Å, *V*=29.020(9) Å$^3$, *Z*=1, $R_w$=0.025, *R*=0.018.

Crystal dimension: 150x150x100μm$^3$, *FW*=45.93, $D_{calc.}$=2.628 g/cm$^3$.

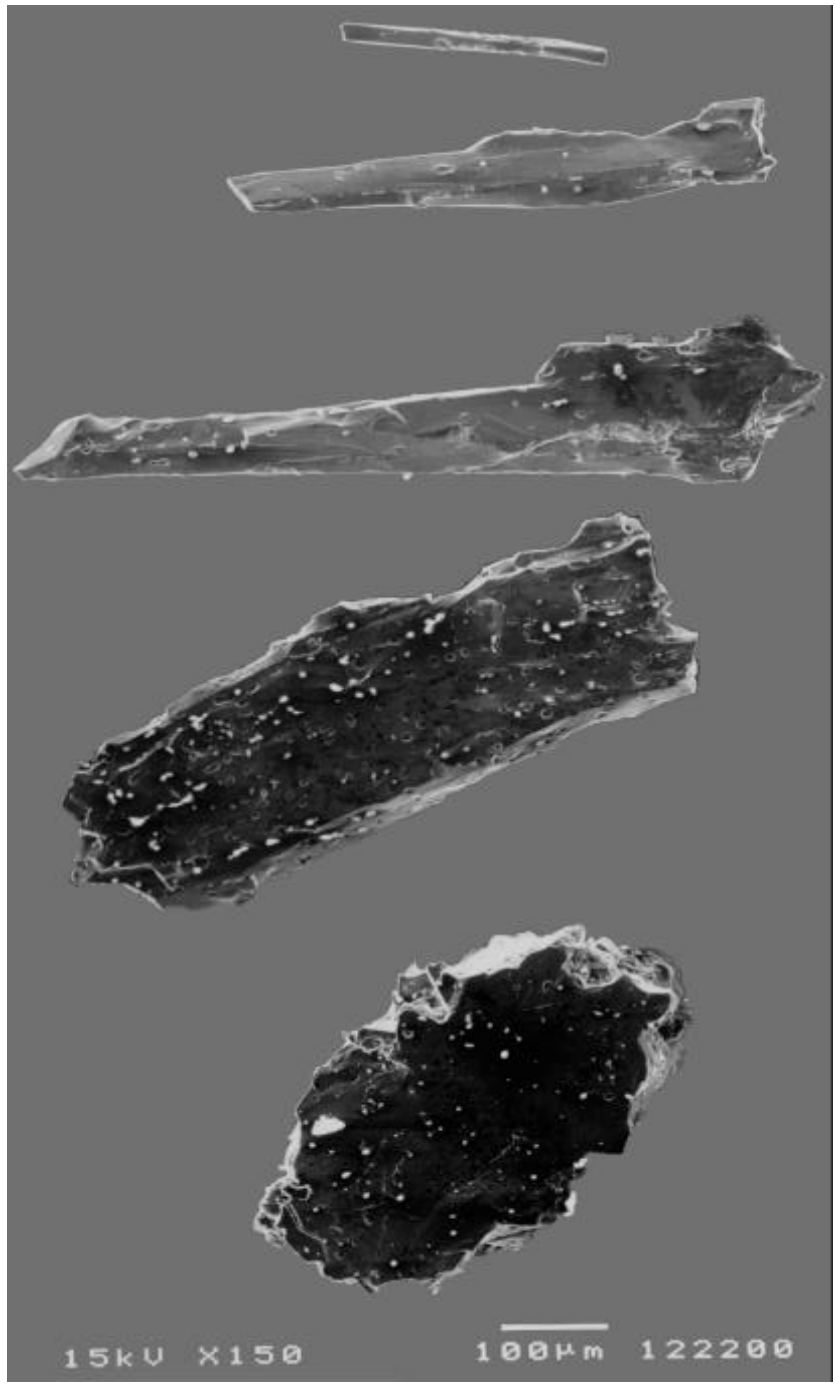

Figure 1

S.Lee et.al.

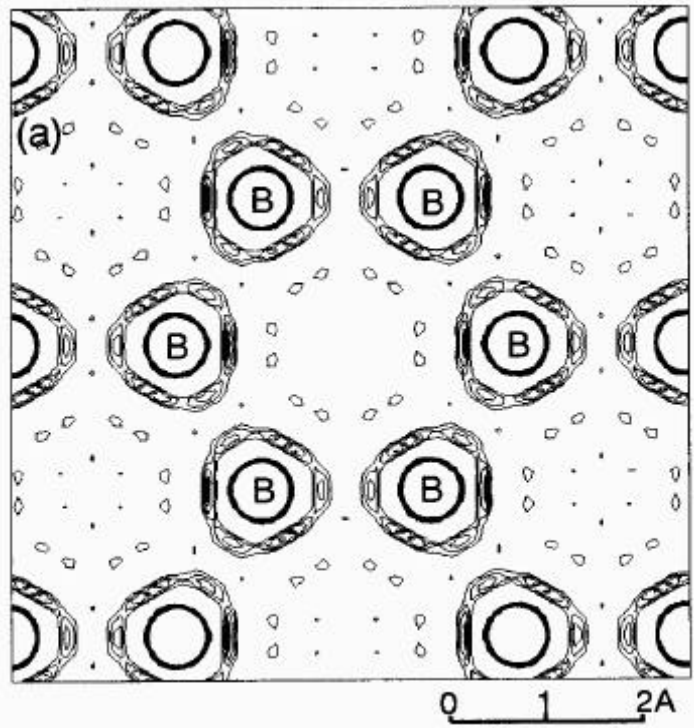

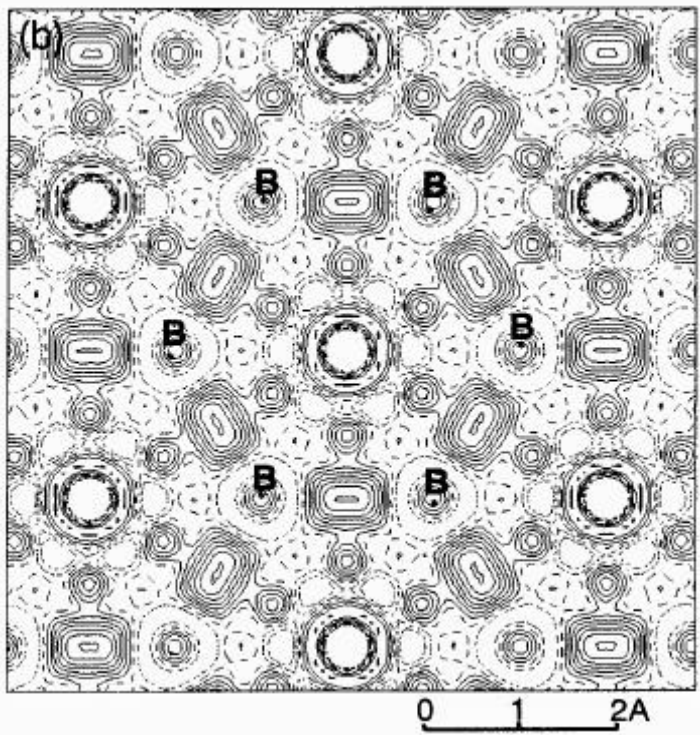

Figure 2

S.Lee et.al.

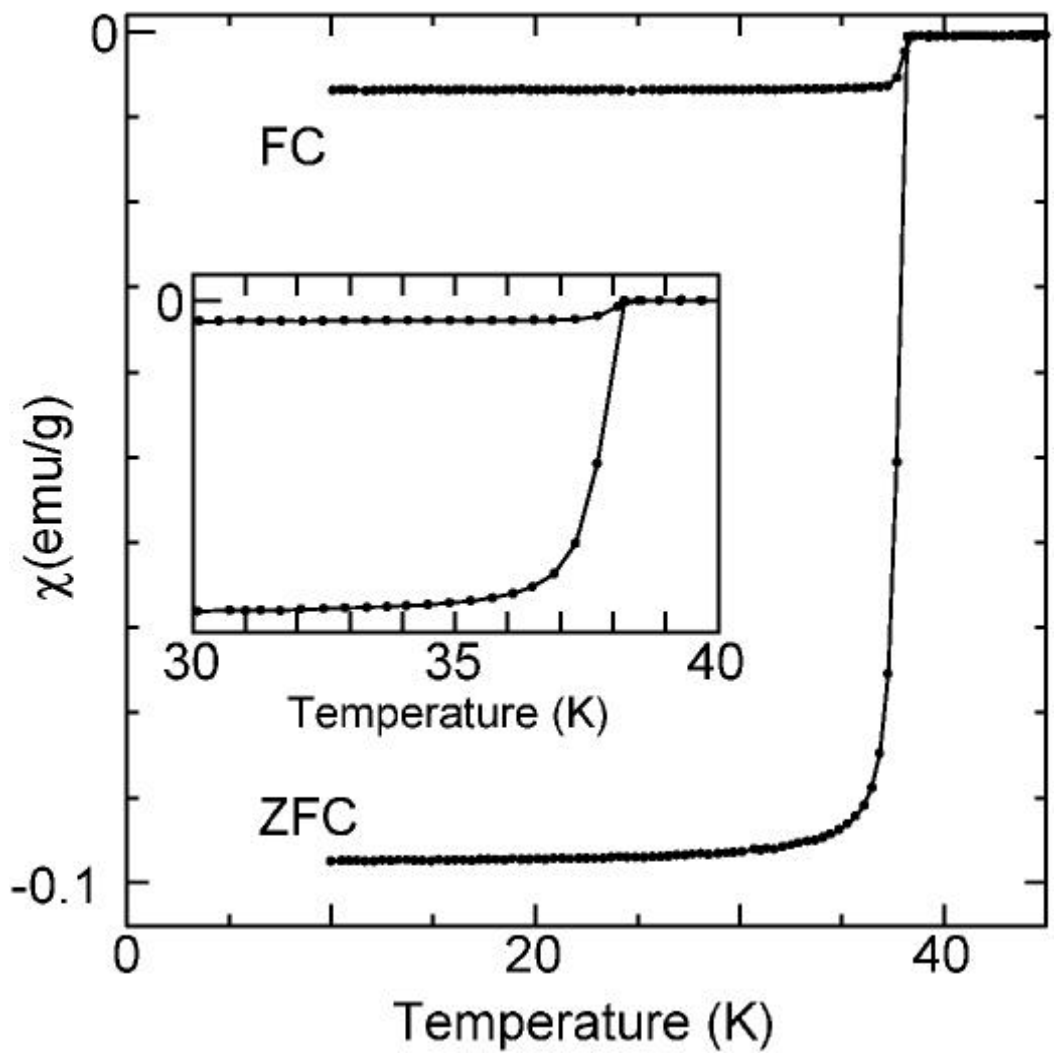

Figure 3

S.Lee et.al.

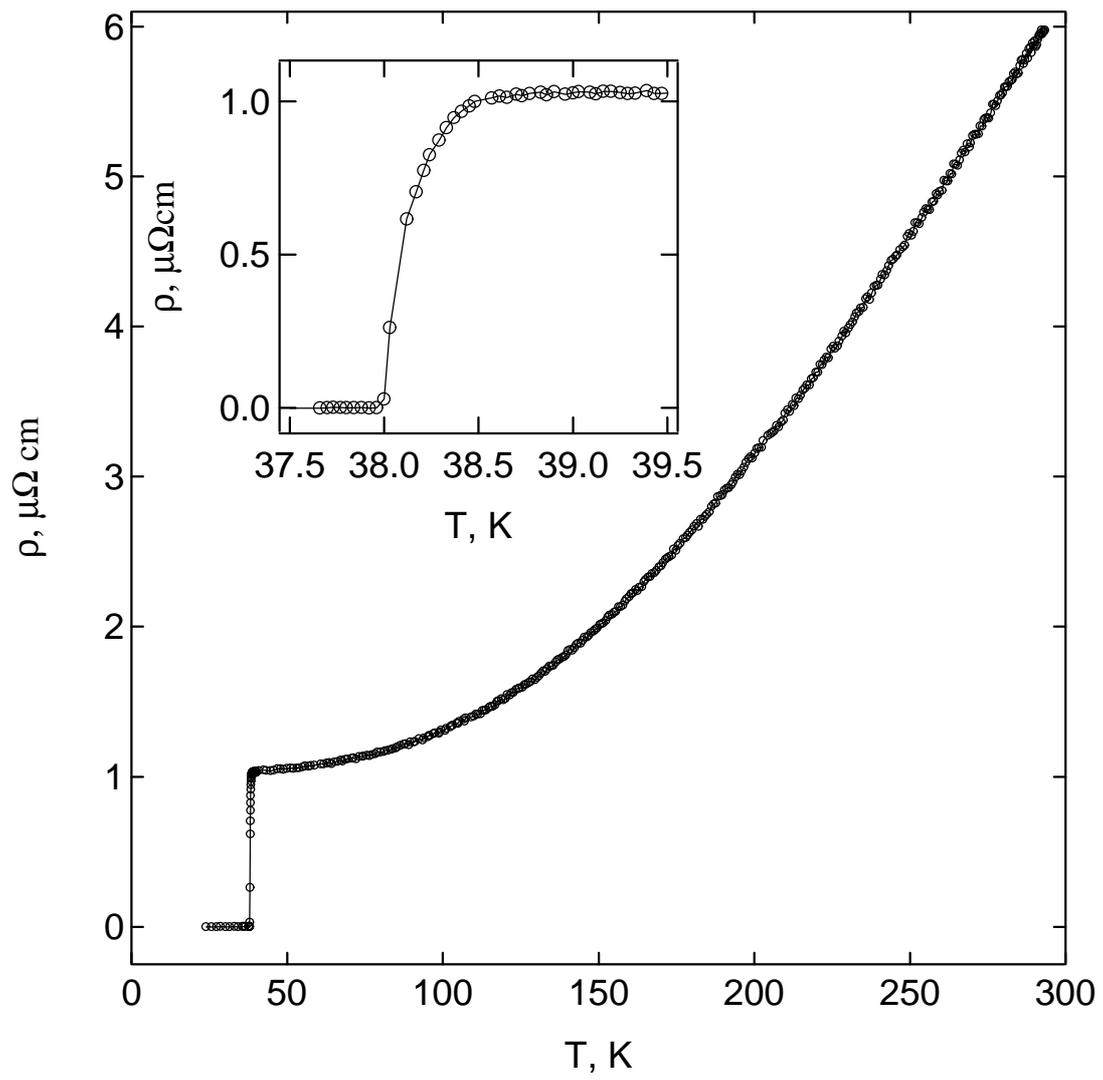

Figure 4

S.Lee et.al.

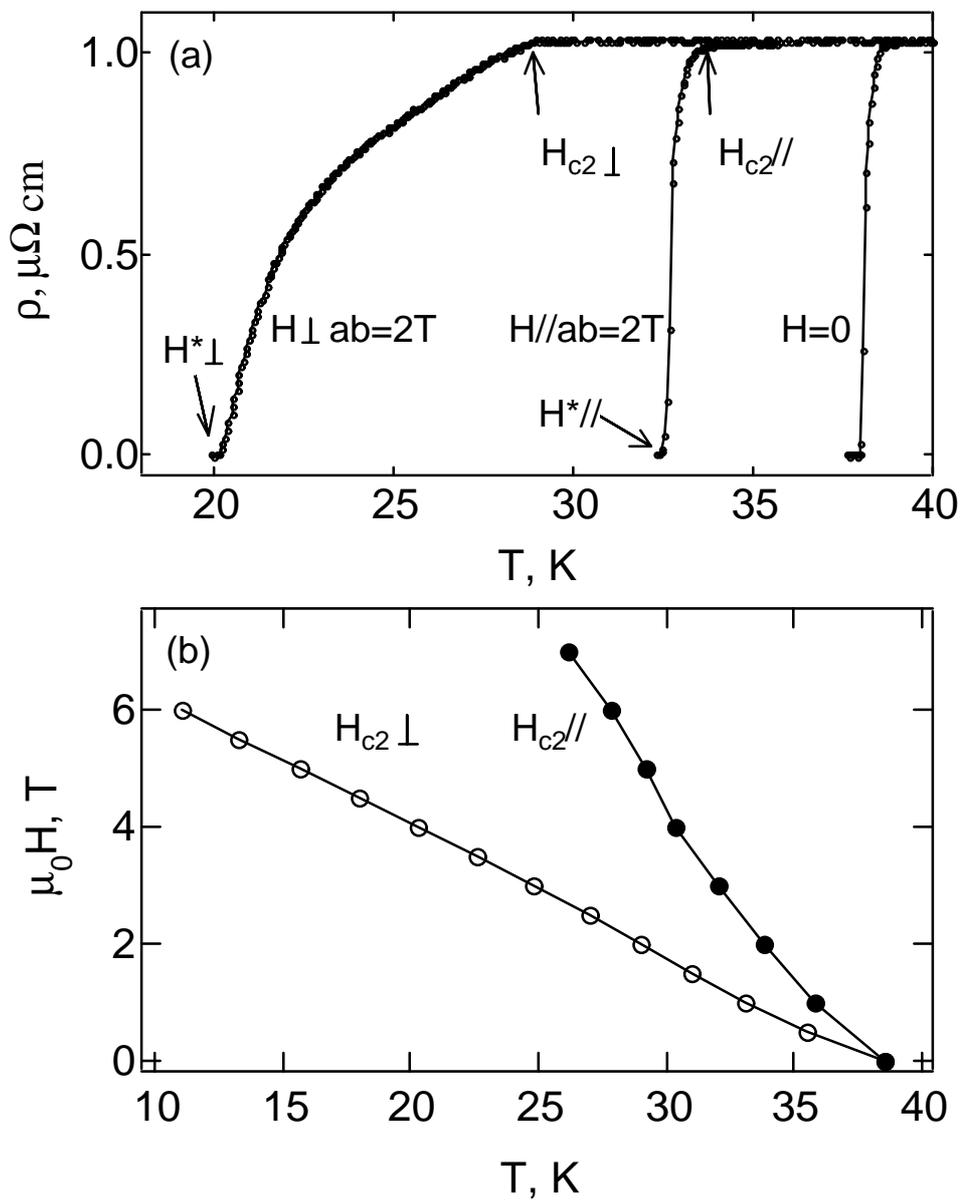

Figure 5

S.Lee et.al.